\documentclass[runningheads]{llncs}
\usepackage[T1]{fontenc} 
\usepackage[utf8]{inputenc} 
\usepackage{amsmath}
\usepackage{amsfonts}
\usepackage{amssymb}
\usepackage{mathtools}
\usepackage{stmaryrd}
\usepackage{bm}
\usepackage{upgreek}
\usepackage{cite}
\usepackage{url}
\usepackage{graphicx}
\usepackage{color}
\usepackage{subfiles}
\usepackage{microtype}
\usepackage{paralist}

\usepackage{tabularx}
\usepackage{booktabs}
\usepackage{subcaption}

\usepackage{siunitx}
\usepackage[square,numbers,sort&compress]{natbib}

\newcommand{\ackedalex}[1]{}

\newif\ifdoubleblind
\doubleblindfalse

\usepackage{xspace}

\title{Uncertainty Estimation in the Real World:\\A Study on Music Emotion Recognition}
\titlerunning{Uncertainty Estimation in Music Emotion Recognition}

\author{%
    Karn N. Watcharasupat\inst{1}\orcidID{0000-0002-3878-5048}
    \and\\
    Yiwei Ding\inst{2,1}\orcidID{0000-0002-8156-3715}
    \and
    T. Aleksandra Ma\inst{1}\orcidID{0009-0001-9254-4106}
    \and\\
    Pavan Seshadri\inst{1}\orcidID{0009-0008-7838-9614}
    \and
    Alexander Lerch\inst{1}\orcidID{0000-0001-6319-578X}%
}
\authorrunning{K. N. Watcharasupat et al.}

\institute{%
Music Informatics Group, Georgia Institute of Technology, Atlanta, GA, USA\\
\email{\{kwatcharasupat, tma98, pseshadri9, alexander.lerch\}@gatech.edu} \and
Center for Artificial Intelligence and Data Science (CAIDAS),\\ University of Würzburg, Würzburg, Germany\\
\email{yiwei.ding@uni-wuerzburg.de}}

\usepackage[bookmarks=false,hidelinks]{hyperref}
\usepackage{cleveref}

\begin{document}

\maketitle
\begin{abstract}
Any data annotation for subjective tasks shows potential variations between individuals. This is particularly true for annotations of emotional responses to musical stimuli. While older approaches to music emotion recognition systems frequently addressed this uncertainty problem through probabilistic modeling, modern systems based on neural networks tend to ignore the variability and focus only on predicting central tendencies of human subjective responses. In this work, we explore several methods for estimating not only the central tendencies of the subjective responses to a musical stimulus, but also for estimating the uncertainty associated with these responses. In particular, we investigate probabilistic loss functions and inference-time random sampling. Experimental results indicate that while the modeling of the central tendencies is achievable, modeling of the uncertainty in subjective responses proves significantly more challenging with currently available approaches even when empirical estimates of variations in the responses are available.

\end{abstract}

\section{Introduction}\label{sec:introduction}
A wide variety of machine-learning tasks require human-annotated data as a proxy for ``ground truths.'' In some tasks, the required annotations are considered mostly ``objective'' (e.g., assigning the label 'cat' to an image), resulting in a  high rate of interrater agreements. In other tasks, the required annotations are inherently subjective, opinion-based, or otherwise ambiguous. This is particularly true in any task involving the use of self-reported psychometric responses to stimuli, where there may not even exist an absolute ``ground truth'' corresponding to a particular stimulus  \cite{Smyth1994InferringGroundTruth}. The ``ground truths'' in these tasks are thus more akin to distributions conditioned on stimulus, that is, associated with some uncertainty. As a result, the corresponding predictions by the model given these stimuli also have associated uncertainties arising from subjective human judgments in the data. Quantifying these uncertainties is crucial as one of the tenets for developing a trustworthy machine learning system. 

\subsubsection{Uncertainty Quantification}
This characterization falls under larger umbrella task Uncertainty Quantification (UQ).
Uncertainties in Machine Learning (ML) systems are commonly categorized into data (aleatoric) uncertainties and model (epistemic) uncertainties \cite{Mucsanyi2024BenchmarkingUncertaintyDisentanglement}. Data uncertainties arise from inherent randomness or noise and are thus considered irreducible. Model uncertainties, on the other hand, depend on the lack of knowledge, {training issues,} or architectural constraints. 

A variety of uncertainty quantification (UQ) methods have been proposed \cite[see][for reviews]{Abdar2021ReviewUncertaintyQuantification, Gawlikowski2023SurveyUncertaintyDeep}. Common methods to characterize model (epistemic) uncertainty include Bayesian methods and ensembling. These methods have been studied, for example, in the context of image classification \cite{Zhang2022ExplainableMachineLearning}, semantic segmentation, biomedical machine learning \cite{Kwon2020UncertaintyQuantificationUsing}, natural language processing \cite{Xiao2019QuantifyingUncertaintiesNatural}, or intersections thereof, with significant efforts dedicated to reducing model uncertainty. Methods to characterize data (aleatoric) uncertainty include direct predictive distribution modeling and deep generative modeling, however, this problem has been historically given somewhat less attention, presumably due to the irreducibility of data uncertainties. A major limitation of many of the existing UQ methods is that they often cannot distinguish between model and data uncertainty, producing a single uncertainty estimate \cite{Mucsanyi2024BenchmarkingUncertaintyDisentanglement}. Model uncertainty estimation, in particular, is often made under the assumption that data uncertainty is negligible. Only in recent years have methods directly quantifying both epistemic and aleatoric uncertainty in a disentangled manner gained some interest. 
It should be noted that most works in this area exclusively focus on classification problems. 

\subsubsection{Music Emotion Recognition}
Music emotion recognition (MER), also known as mood recognition, is an archetypal example of a task without an absolute ground truth. The recognition of emotion in music is a particularly important problem in music information retrieval (MIR), as emotional cues are commonly used as descriptors for a piece of music. The annotations corresponding to a particular stimulus are usually either categorical affective labels (e.g., relaxing, suspenseful) or numerical scores corresponding to different levels of various ``dimensions'' of emotions (e.g., Likert ratings on valence and arousal, see below) \cite[pp. 15--20]{Yang2011MusicEmotionRecognition}. 

Depending on the type of annotations, MER can be seen as either a classification or a regression task. A common formation of regression-based MER involves associating a stimulus to a point in a low-dimensional ``space'' of emotion, with the most common being the valence-arousal (V-A) model and variations thereof \cite[see][pp. 55--88]{Yang2011MusicEmotionRecognition}. The V-A model is a simplified two-dimensional representation of emotion, originating from Russell's circumplex model of affect \cite{Russell1980CircumplexModelAffect}. The \textit{valence} dimension accounts for ``the degree to which an emotion is
associated with favorable outcomes,'' while the \textit{arousal} dimension accounts for ``the amount of
energy associated with an emotion''
\cite[p. 245]{Tan2018PsychologyMusicSound}. The remainder of this work will focus on MER as a multiple regression problem in the V-A space.

There is no absolute nor objective ground truth available for such an MER task since 
\begin{inparaenum}[(i)]
    \item different individuals might experience the emotional content and impact of the same piece of music differently \cite[p. 107]{Yang2011MusicEmotionRecognition}, and
    \item two  similar affective reactions may be ascribed to two very different numerical values during data collection \cite{Lionello2021IntroducingMethodIntervals}.
\end{inparaenum}

Interestingly, while the distributional nature of psychometric data has been accounted for in several older feature-based approaches to MER via probabilistic modeling \cite{Schmidt2010PredictionTimevaryingMusical, Schmidt2010PredictionTimeVaryingMusicala, Schmidt2011ModelingMusicalEmotion, Yang2011PredictionDistributionPerceived, Wang2012AcousticEmotionGaussians, Imbrasaite2013EmotionTrackingMusic, Imbrasaite2014CCNFContinuousEmotion, Wang2015ModelingAffectiveContent, Wang2015HistogramDensityModeling, Chen2017ComponentTyingMixture, Chin2018PredictingProbabilityDensity}, most modern neural MER systems do not take into account the distributional information of these empirical annotations. Instead, central tendencies such as the mean and median of the ratings, are solely predicted without acknowledging that these are ultimately sample estimates of random variables from unknown distributions. As most building blocks of deep learning systems have been designed as deterministic maps, they are arguably not always well suited for psychometric targets where variations in the ground truth cannot be simply treated as label noise. 

In recent years, MER has started to become utilized beyond purely media and entertainment applications, particularly in music therapy for both psychiatric and non-psychiatric treatments \cite{Cui2022ReviewMusicemotionRecognition}, where ML trustworthiness is of crucial importance. 
As a result, our study\footnote{Code and other supplementary materials are available at the anonymized repository \href{https://anonymous.4open.science/r/emotionally-uncertain}{https://anonymous.4open.science/r/emotionally-uncertain}.} aims to investigate the extent to which these stochasticities, specifically, interrater variations, can be modeled with a deep learning approach, acknowledging that current MER systems are utilizing deep neural architectures. In order to do so, we explore several methods in which such a system can learn to predict a probability distribution or an empirical approximation thereof.
To summarize, this study
\begin{inparaenum}[(i)]
    \item identifies approaches to  model the uncertainty in rater annotations of highly subjective regression tasks, and
    \item explores and benchmarks these approaches in the context of music emotion recognition.
\end{inparaenum}





\section{Methods}\label{sec:methods}

\begin{table}
    \centering
    \footnotesize
    \setlength{\tabcolsep}{3pt}
    \begin{tabularx}{\columnwidth}{Xcccccc}
        \toprule 
        & \multicolumn{2}{c}{Training Targets} & \multicolumn{2}{c}{Outputs} & \multicolumn{2}{c}{Requires Multiple Runs?}\\
        \cmidrule(lr){2-3}\cmidrule(lr){4-5}\cmidrule(lr){6-7}
        Method &  Raw & Interpreted & Raw & Interpreted & Train & Inference\\
        \midrule
        Seeds & $\mu$ & $\mu$ &$\{\hat{y}_i\}$ & $\mathcal{N}(\langle \hat{y} \rangle, s_{\text{seed}}^2)$ & Y & Y \\[3pt]
        MC Dropout & $\mu$ &$\mu$ & $\{\Tilde{y}_i\}$ & $\mathcal{N}(\langle \Tilde{y} \rangle, s_{\text{MC}}^2)$  & N & Y \\[3pt]
        NLL & $\mu$ &$\mu$ & $\hat{\mu}, \hat{\sigma}$ & $\mathcal{N}(\hat{\mu}, \hat{\sigma}^2)$ &  N & N \\
        \midrule
        MSE & $\mu, \sigma$ & $\mu, \sigma$ & $\hat{\mu}, \hat{\sigma}$ & $\hat{\mu}, \hat{\sigma}$ & N & N \\
        KLD & $\mu, \sigma$ & $\mathcal{N}(\mu, \sigma^2)$ & $\hat{\mu}, \hat{\sigma}$ & $\mathcal{N}(\hat{\mu}, \hat{\sigma}^2)$ & N & N \\
        \bottomrule
    \end{tabularx}
    
    \caption{Training target(s), output(s), their interpretations, and multiple-run requirements for the methods tested.}
    \label{tab:methods}
\end{table}

In this section, we first outline the statistical preliminaries and then describe in detail the different methods employed to estimate uncertainty. As noted earlier, these methods cannot strictly distinguish data and model uncertainties from each other. Although we are more interested in modeling data uncertainty, we also investigate methods for estimating model uncertainty for comparison.
Practically, we simply categorize the methods by whether they require ground-truth uncertainty during training or not.
We assume~---~where necessary~---~that both ground-truth data distribution and output distribution are Gaussian.
The assumptions and properties of all investigated methods are listed in Table~\ref{tab:methods}. These methods are chosen for their common usage in literature as baselines and for their ease of interpretation. Many of the state-of-the-art methods either only work for classification tasks or require invasive changes to the models, rendering them impractical for a regression task with low data availability such as MER. 

\subsection{Preliminaries}

\begin{figure}
    \centering
    \begin{subfigure}[b]{0.3\textwidth}
        \includegraphics[width=\textwidth]{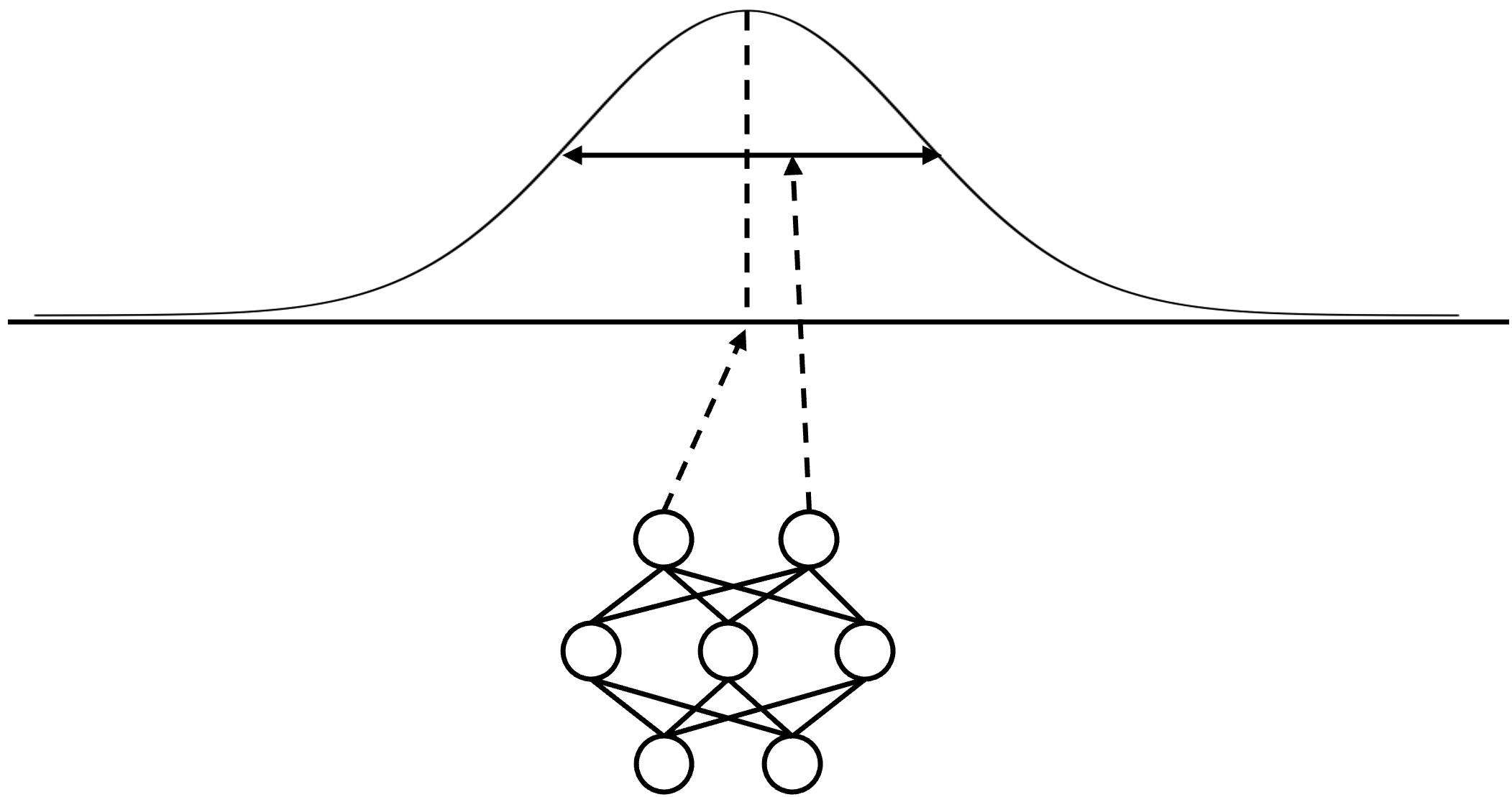}
        \caption{Direct estimation of SD}
        \label{fig:prob_loss}
    \end{subfigure}
    \begin{subfigure}[b]{0.3\textwidth}
        \includegraphics[width=\textwidth]{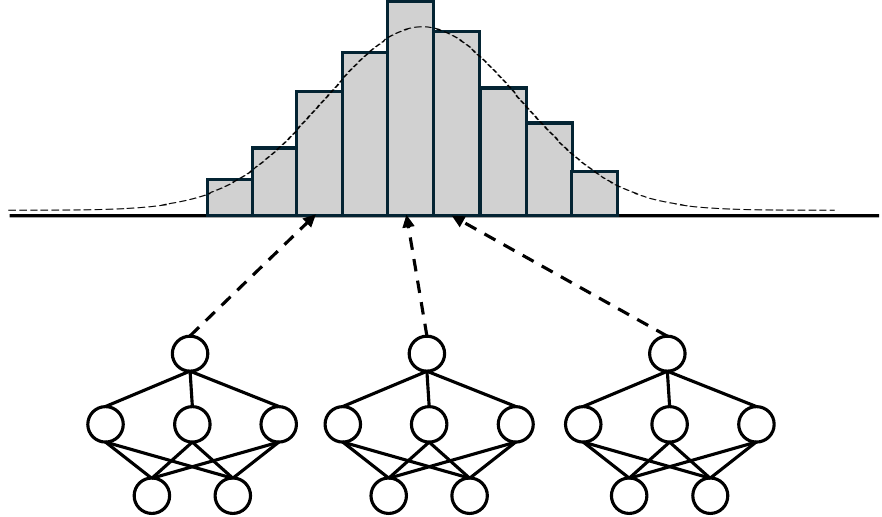}
        \caption{Random seeds}
        \label{fig:seeds}
    \end{subfigure}
    \begin{subfigure}[b]{0.3\textwidth}
        \includegraphics[width=\textwidth]{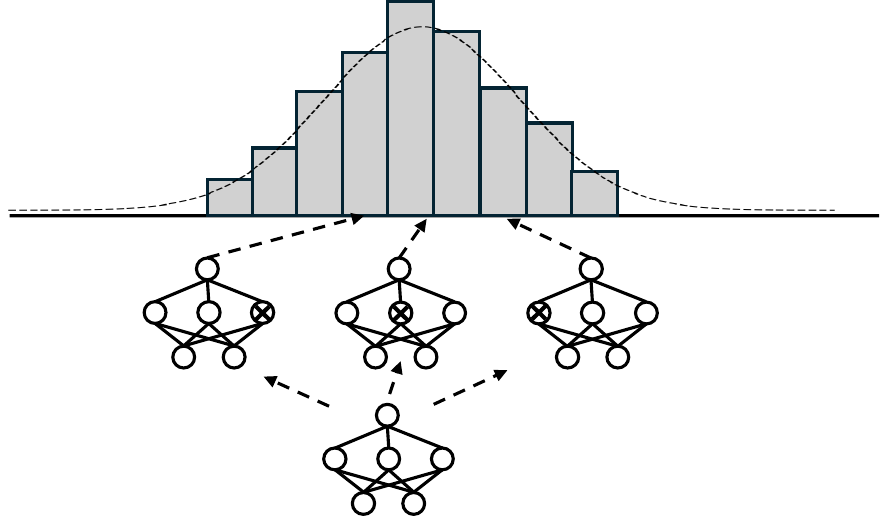}
        \caption{Monte Carlo dropout}
        \label{fig:mc_dropout}
    \end{subfigure}
    \caption{Illustration of different ways of uncertainty estimation.}
    \label{fig:uncertainty}
\end{figure}

Following previous work \cite{Schmidt2010PredictionTimevaryingMusical}, we use valence and arousal to measure the emotion of music. Formally, we model an emotional response to a stimulus $x$ as a bivariate conditional variable $Y \mid X = x \sim \mathcal{N}(\bm{\upmu}_{x}, \mathbf{\Sigma}_x)$. Unlike Schmidt and Kim \cite{Schmidt2010PredictionTimevaryingMusical}, however, we model $\mathbf{\Sigma}_x$ as a diagonal matrix. This assumption is taken to simplify the downstream methodology and analyses, especially as the prediction of a positive semi-definite matrix in deep learning is nontrivial and prone to numerical instability. Equivalently, this model can be composed into two independent random variables: valence $V \mid x \sim \mathcal{N}(\mu_{V, x}, \sigma^2_{V, x})$ and arousal $A \mid x \sim \mathcal{N}(\mu_{A, x}, \sigma^2_{A, x})$. Consequently, the spread parameter $\sigma_{\cdot, x}$ models the level of uncertainty of the emotional responses by individuals for a particular stimulus $x$. The sources of variations accounted for in an empirical estimate of $\sigma_{\cdot, x}$ could include the individual demographics, the listening environment, and familiarity with the musical genre, amongst others \cite{Yang2007MusicEmotionRecognition}. For brevity, we will use $\mu$ and $\sigma$ in the following to denote \textit{empirical} estimates of the mean and standard deviation in subsequent sections. Figure~\ref{fig:uncertainty} visualizes the different concepts of uncertainty estimation introduced below.

\subsection{Methods requiring empirical uncertainty}

In this section, we present training methods for joint mean-uncertainty estimation with the data requirement of empirical uncertainty as training targets, in addition to the V-A ratings.
In both methods below, the model outputs estimates of the distributional parameters $(\hat{y}, \hat{\sigma}) \coloneqq f(x \mid \hat{\theta})$.

\subsubsection{MSE Loss:} The first method naively aims to minimize the mean square error (MSE) on the distributional parameters by minimizing the loss function whose sample contribution is given by 
\begin{align}
    \mathcal{L}_{\mathrm{MSE}}(\hat{\mu}, \hat{\sigma}; \mu, \sigma) = (\hat{\mu} - \mu)^2 + (\hat{\sigma}^2 - \sigma^2)^2\,.
\end{align}

\subsubsection{KLD Loss:} The second method requires probabilistic modeling of the output as a Gaussian distribution. In doing so, it is possible to minimize the KL divergence (KLD) between the predicted and the empirical distributions $\mathbb{D}_{\mathrm{KL}}\{\mathcal{N}(\hat{\mu}, \hat{\sigma}^2)\ \|\ \mathcal{N}({\mu}, {\sigma}^2)\}$, via a mini-batch proxy
\begin{align}
    \mathcal{L}_{\mathrm{KLD}}(\hat{\mu}, \hat{\sigma}; \mu, \sigma) &= \dfrac{1}{2}\left(\dfrac{\hat{\mu} - \mu}{\hat{\sigma}}\right)^2 - \log \hat{\sigma} + \dfrac{1}{2}\dfrac{\hat{\sigma}^2}{\sigma^2}.
\end{align}
The uncertainty estimate in both cases is directly the predicted SD $\hat{\sigma}$.

\subsection{Methods not requiring empirical uncertainty}

Given that multi-annotator datasets with a sufficient number of raters are relatively rare in MER, we also consider methods where empirical estimates of the rating uncertainty are not needed during training.

\subsubsection{NLL Loss:}
Following \cite{Ooi2022ProbablyPleasantNeuralProbabilistic, Watcharasupat2022AutonomousInSituSoundscape, Ooi2023AutonomousSoundscapeAugmentation}, probabilistic modeling of the outputs can still be employed even when ground truth uncertainty is unavailable. This is done by requiring a two-output model as per above and viewing the estimates $\hat{Y} \mid X$ as a Gaussian random variable parametrized by the outputs of $f(X \mid \hat{\theta})$, where $\hat{\theta}$ is the learnt model parameters. By using a mini-batch proxy for minimizing $\mathbb{E}_{\mathfrak{D}}\{(Y - f(X \mid \hat{\theta}))^2\}$ over a dataset $\mathfrak{D}$, it is possible to arrive the negative log-likelihood loss (NLL) contributions as a practical proxy, where
\begin{align}
\label{eq:nll}
    \mathcal{L}_{\text{NLL}}(\hat{\mu}, \hat{\sigma}; \mu) &= \dfrac{1}{2}\left(\dfrac{\hat{\mu} - \mu}{\hat{\sigma}}\right)^2 + \dfrac{1}{2}\log \hat{\sigma}\,.
\end{align}
As with the KLD method, the uncertainty estimate in both cases is directly the predicted SD $\hat{\sigma}$. The same loss and variations thereof have also been used in other UQ studies \cite{Seitzer2022PitfallsHeteroscedasticUncertainty, Valdenegro-Toro2022DeeperLookAleatoric}, given its direct relation to maximum likelihood estimation.



Clearly, the NLL loss function has some similarity with the KLD loss function. Both are effectively some forms of a pseudo-regularized generalization of the MSE, with the difference lying in the pseudo-regularizer. In the NLL loss, the first term leans towards bigger $\hat{\sigma}^2$, especially for samples that have larger MSE, while the second term leans towards lower $\hat{\sigma}^2$, which encourages the model to be more ``certain'' about the predictions. 



\subsubsection{Random Seeds:}
Regardless of the loss function, using $n$ different random seeds during the training runs can be seen as drawing independent and identically distributed (i.i.d.) samples of optimal parameters from a conditional distribution:
\begin{align*}
    \hat{\theta}_1, \hat{\theta}_2, ..., \hat{\theta}_n \stackrel{\mathrm{i.i.d.}}{\sim} \mathbb{P}_{\Theta \mid \mathfrak{D}}\,.
\end{align*}
where $\mathfrak{D}$ is the dataset. Let  $\hat{y}_i \coloneqq f(x \mid \hat{\theta}_i)$ as the rating estimate from the model $\hat{\theta}_i$ with the $i$th random seed. 
For each input $x$, the averaged prediction over seed is thus given by 
\begin{align}
    \langle{\hat{y}_i}\rangle = \dfrac{1}{n}\sum_{i=1}^n \hat{y}_i \approxeq \mathbb{E}_{\Theta \mid \mathfrak{D}}\{f(x \mid \Theta)\}
\end{align} 
while the uncertainty estimate is given by
\begin{align}
    s_{\text{seed}}^2 (x) = \dfrac{1}{n-1}\sum_{i=1}^n(\hat{y}_i - \langle{\hat{y}_i}\rangle)^2 \,.
\end{align}
The benefit of using random seeds for uncertainty estimation lies in that there is no need for the model to explicitly output an uncertainty estimate. However, this method requires a considerable amount of training runs, thus resultant models, to obtain a relatively accurate estimation of the uncertainty. Therefore, this method requires a relatively large amount of both computational and storage resources compared to other methods. Additionally, this method is more commonly used to estimate model uncertainty rather than data uncertainty. This method is included to investigate whether variations in the model optima, thus the variations in the prediction for each stimulus, have any relation to the stimulus-dependent variations.

\subsubsection{MC Dropout:}

Given that the multiple-seed method requires a significant number of model instantiations, another~---more efficient---~possibility of obtaining multiple predictions from a single model is to turn on dropout during inference (``Monte Carlo dropout''). Following \cite{Gal2016BayesianConvolutionalNeural, Kendall2017WhatUncertaintiesWe}, in a system with dropout layers, turning on dropout during inference can be seen as a simulation of drawing i.i.d. parameter samples from a distribution. 
By repeating the process multiple times using an already optimized set of parameters $\hat{\theta}$, we have
\begin{align*}
    \Tilde{\theta}_1, \Tilde{\theta}_2, ..., \Tilde{\theta}_n \mathop{\sim}\limits^{\mathrm{i.i.d.}} \mathbb{P}_{\Tilde{\Theta} \mid \hat{\theta}, \mathfrak{D}} \,,
\end{align*}
where $\Tilde{\theta}$ is the parameters during a single inference run. 
Similar to the Random Seeds method, the prediction for input $x$ is 
\begin{align}
    \langle{\tilde{y}_i}\rangle = \dfrac{1}{n}\sum_{i=1}^n \tilde{y}_i \approxeq \mathbb{E}_{\tilde{\Theta} \mid \hat{\theta}, \mathfrak{D}}\{f(x \mid \tilde{\Theta})\}
\end{align} 
while the uncertainty estimate is given by
\begin{align}
    s^2_{\text{MC}} (x) = \dfrac{1}{n-1}\sum_{i=1}^n(\Tilde{y}_i - \langle\Tilde{y}_i\rangle)^2\,.
\end{align}
Compared to using different seeds, MC Dropout requires only one training run, significantly reducing the training cost.
As in the previous method, this method is more often used to estimate model uncertainty.

\section{Experimental Setup}\label{sec:experimental}
In this section, we describe our experimental setup, including the dataset and the model used.

\subsection{Data}
\begin{figure}[t]
    \centering
    \begin{minipage}[t]{0.47\textwidth}
        \centering
        \includegraphics[width=2.5in]{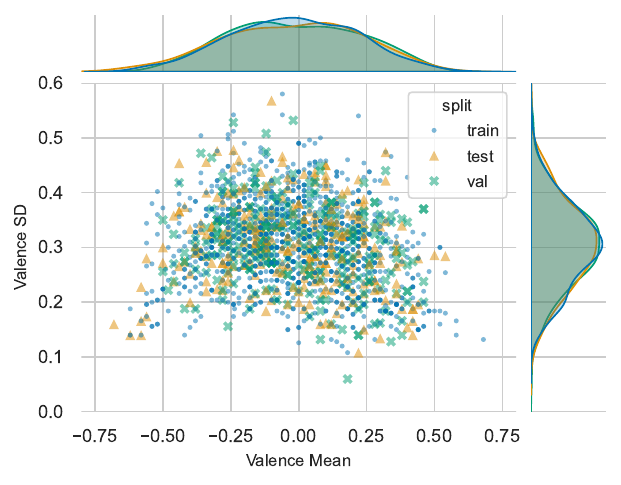} 
        \caption{Distribution of mean and SD of valence ratings}
        \label{fig:eda-valence}
    \end{minipage}
    \hfill
    \begin{minipage}[t]{0.47\textwidth}
        \centering
        \includegraphics[width=2.5in]{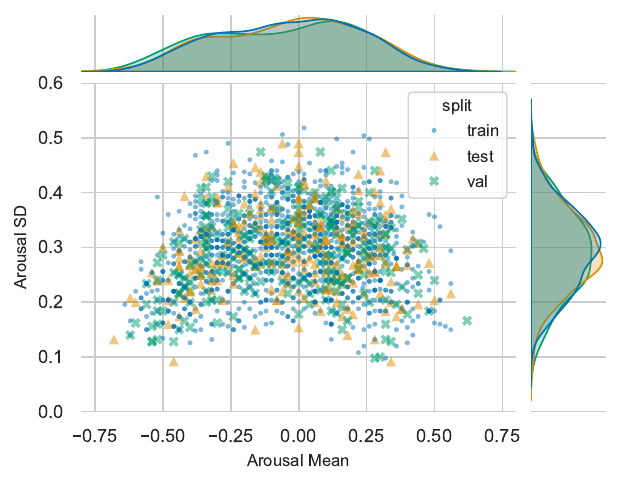} 
        \caption{Distribution of mean and SD of arousal ratings}
        \label{fig:eda-arousal}
    \end{minipage}
\end{figure}

The dataset used for this study, the MediaEval Database for Emotional Analysis in Music (DEAM) dataset \cite{Aljanaki2017DevelopingBenchmarkEmotional}, consists of 1802 songs. In this work, we will only use a subset of 45-second excerpts from 1744 songs whose annotations were collected using a static setup in 2013 and 2014, with at least 10 annotators per song. The remaining 58 songs collected in 2015 with 5 annotators using a dynamic setup were not used due to the smaller number of annotators and the difference in data collection methodology compared to the previous collection efforts.

In the original 2014 MediaEval setup, the 744 songs collected in 2013 were used as the training set, and the 1000 songs collected in 2014 were used as the evaluation set. In this work, we restructured the dataset into train-validation-test splits with a 70:15:15 ratio, stratified by genres. The song identifiers for each split are published in the accompanying repository for reproducibility. \Cref{fig:eda-valence,fig:eda-arousal}  show the distribution of the mean and SD of the normalized valence and arousal, respectively, aggregated across all raters of a given song. The normalization is discussed below. Each point in the scatter plot represents a song in the dataset, differentiated in color by the data split. The marginal density plots represent the distribution of the mean and SD respectively. The density curves for the split have been normalized to compare their distribution. As shown in the marginal distribution plots in \Cref{fig:eda-valence,fig:eda-arousal}, the distributions of the mean and SD of both the valence and the arousal ratings are also approximately Gaussian. We note that almost half of the standard deviations are greater than 0.3 and can be as large as 0.5. Given that most mean values are between -0.5 and 0.5, the inter-rater variations can be considerable, and therefore this uncertainty should be non-ignorable in this dataset.

\subsection{Data Preprocessing}
The data collection for DEAM follows a 9-point Likert scale from 1 to 9, with $r_{\text{neutral}} = 5$ being the neutral point, leaving $R=4$ non-neutral options on each end. 
The ratings were normalized with 
\begin{align}
    h(r) &= \dfrac{r-r_{\text{neutral}}}{R+1}
\end{align}
to bring them to the $[{-1+\delta}, {1-\delta}] \subset (-1, 1)$ range, $\delta = 1/(R+1)$. This normalization is similar in motivation to label smoothing \cite{Muller2019WhenDoesLabel}. It was adopted so that the two most extreme ratings can be achieved non-asymptotically, without requiring disproportionately large logit values before a sigmoidal activation. 

\subsection{Model}

The model used in this work is a combination of the MusicFM-MSD model \cite{Won2024FoundationModelMusic} and a smaller fully connected network (FCN) with 2 hidden layers. Each hidden FCN layer consists of 128 hidden neurons, with ELU activation and \SI{50}{\percent} dropout. 
The output layer has one output for the mean-only model and two outputs for the mean-variance model. The mean head has hyperbolic tangent activation while the variance head has a negative softplus activation given by
\begin{align}
    \operatorname{NegSoftPlus}(z) = - \log(1+\exp z)) \in (-\infty, 0)
\end{align}
for stability. In effect, we have $\hat{\sigma} = (1+e^{z})^{-1} \in (0, 1)$ for a corresponding finite-valued logit $z$. The architecture of the smaller network was chosen via preliminary experiments to yield a model with a sufficient complexity that does not overfit in the relatively small MER dataset, especially since data augmentation cannot be easily performed for MER. 

Each stereo audio signal from DEAM was first downsampled to \SI{24}{\kilo\hertz}, then padded or trimmed to \SI{45}{\second}. The preprocessed audio signal was then passed through the foundation model to obtain a 1024-dimensional time series of features for each audio channel. The features were averaged over audio channels and time frames to obtain a single 1024-dimensional vector per song, which was then passed through the FCN to obtain the predictions.

\subsection{Training}

For all experiments in this work, the foundation model was frozen. The FCN model was trained with an Adam optimizer with an initial learning rate of \num{e-3} for up to \num{100} epochs. The learning rate is dropped by a factor of \num{0.9} after 3 epochs of no improvement in validation loss, subject to a minimum learning rate of \num{e-5}. Each epoch consists of 128 training batches of size 32 each, totaling 4096 training samples per epoch. For each training sample, we take the mean (and standard deviation if required) of all the raters as the ground-truth.


\section{Results and Discussions}\label{sec:results}
\begin{figure*}[t]
    \centering
    \includegraphics[width=\linewidth]{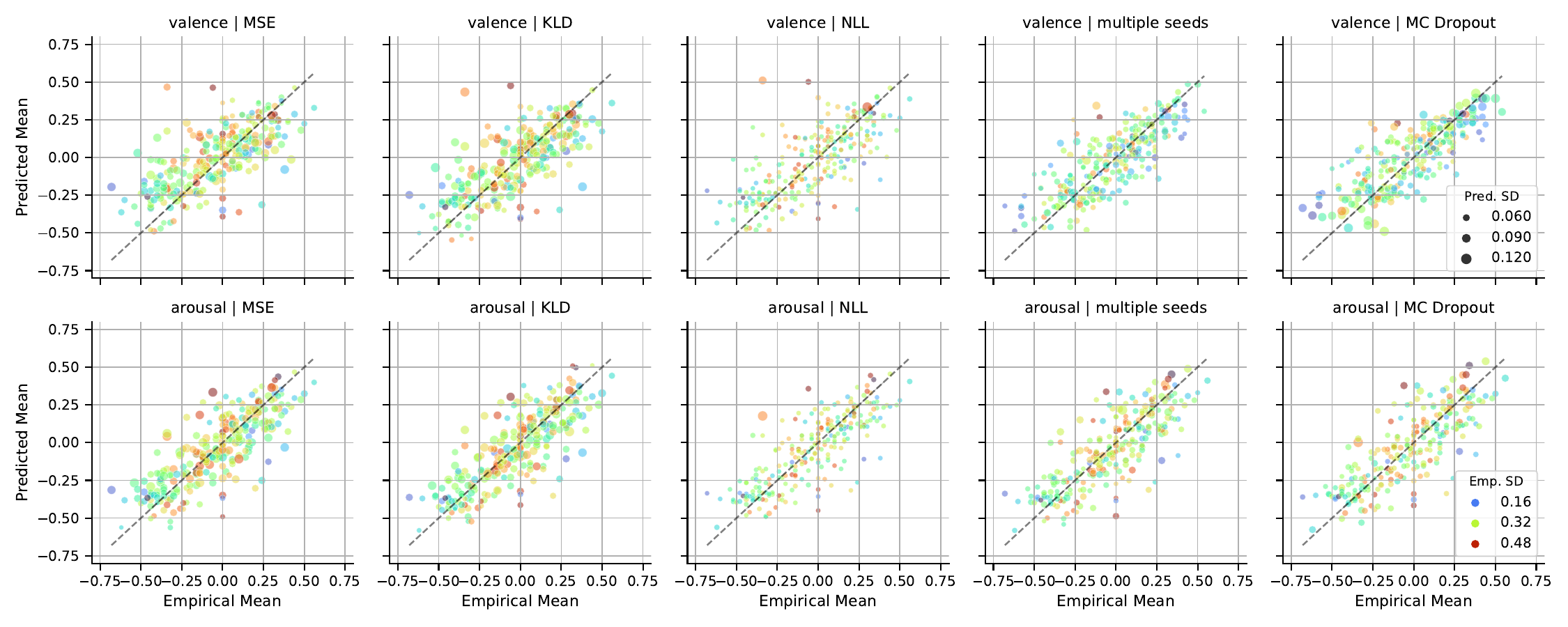}
    \caption{Empirical and corresponding predicted means of arousal and valence. For single-model methods, the visualization was derived from the outputs of one model realization (seed 41).}
    \label{fig:mean-summary}
\end{figure*}

\begin{table*}[t]

\renewrobustcmd{\bfseries}{\fontseries{b}\selectfont}
\renewrobustcmd{\boldmath}{}
    \centering
    \scriptsize
    \begin{tabularx}{\linewidth}{X *{6}{S[table-format=-.2(.2),separate-uncertainty=true,multi-part-units=single]}}
        \toprule
                            & \multicolumn{6}{c}{Mean}\\
        \cmidrule(lr){2-7}
                            & \multicolumn{2}{c}{$R^2$}     & \multicolumn{2}{c}{Pearson $r_p$}     & \multicolumn{2}{c}{Spearman $r_s$}\\
        \cmidrule(lr){2-3}\cmidrule(lr){4-5}\cmidrule(lr){6-7}
                            & {Valence} & {Arousal} & {Valence} & {Arousal}& {Valence} & {Arousal}\\
        \midrule
         Random Seeds   & 0.59(0.04) &\bfseries 0.62(0.02) & 0.78(0.02) & 0.80(0.01) & 0.78(0.02) & \bfseries0.80(0.01)\\
         NLL Loss           & \bfseries 0.61(0.03) & 0.61(0.02) & \bfseries0.79(0.02) & 0.80(0.01) & \bfseries0.79(0.02) & \bfseries0.80(0.01) \\
         \midrule
         MSE Loss           & 0.59(0.02) & \bfseries0.62(0.01) & 0.78(0.02) & 0.80(0.01) & 0.78(0.02) & \bfseries0.80(0.01) \\
         KLD Loss        & 0.55(0.02) & \bfseries0.62(0.02) & 0.76(0.01) & \bfseries0.81(0.01) & 0.76(0.01) & \bfseries0.80(0.01) \\
         \midrule
         TNN-SVR \cite{Cheuk2020RegressionbasedMusicEmotion}
         &  0.67 & 0.36 \\
         \bottomrule
    \end{tabularx}
    \caption{Metrics and their standard deviations for predictions of the mean subjective ratings by different methods, aggregated over 15 random seeds. Note that the results from TNN-SVR were obtained from literature with different data splits.}
    \label{tab:results_mean}
\end{table*}

\begin{figure*}[t]
    \centering
    \includegraphics[width=\linewidth]{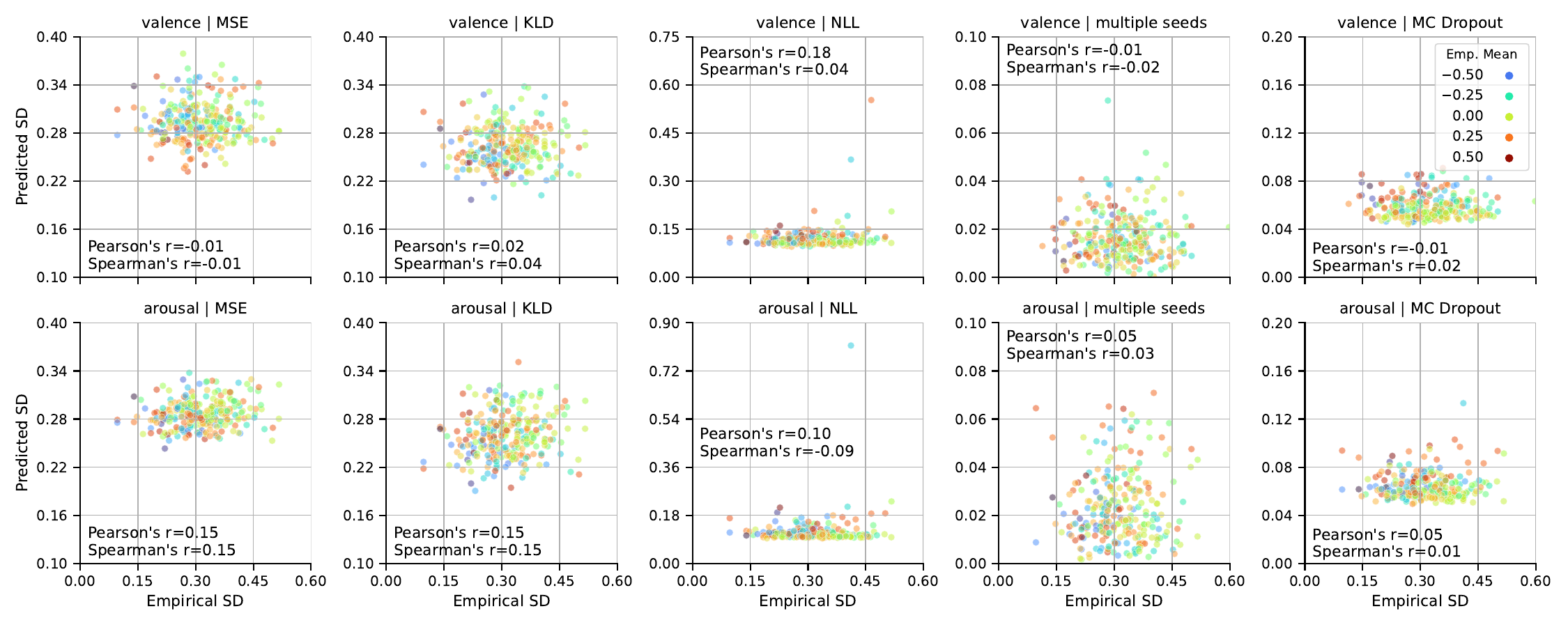}
    \caption{Empirical and corresponding predicted standard deviations of arousal and valence. For single-model methods, the visualization was derived from the outputs of one model realization (seed 41).}
    \label{fig:std-summary}

    
\end{figure*}

\begin{table*}[t]

\renewrobustcmd{\bfseries}{\fontseries{b}\selectfont}
\renewrobustcmd{\boldmath}{}
    \centering
    \scriptsize
    \begin{tabularx}{\linewidth}{X cc *{4}{S[table-format=-0.2(0.2),separate-uncertainty=true,table-align-uncertainty=true,multi-part-units=single]}}
        \toprule
                            & \multicolumn{6}{c}{Standard Deviation}\\
        \cmidrule(lr){2-7}
                            & \multicolumn{2}{c}{$R^2$}     & \multicolumn{2}{c}{Pearson $r_p$}     & \multicolumn{2}{c}{Spearman $r_s$}\\
        \cmidrule(lr){2-3}\cmidrule(lr){4-5}\cmidrule(lr){6-7}
                            & {Valence} & {Arousal}& {Valence} & {Arousal}& {Valence} & {Arousal}\\
        \midrule
         Random Seeds       & {$\ll$0}        & {$\ll$0}    & -0.06 & \bfseries 0.07 & -0.05 & \bfseries 0.08\\ 
         MC Dropout         & {$\ll$0}        & {$\ll$0}    & \bfseries 0.01(0.04) & -0.01(0.05) & \bfseries  0.03(0.03) & -0.07(0.05) \\
        NLL Loss            & {$\ll$0} & {$\ll$0} &-0.07(0.03) & -0.08(0.02) & -0.02(0.05) & 0.03(0.04) \\ \midrule
        MSE Loss           & {$\ll$0} & {$\ll$0}  & -0.16(0.05) & -0.15(0.02) & -0.16(0.05) & -0.14(0.03) \\
        KLD Loss           & {$\ll$0} & {$\ll$0} & -0.12(0.03) & -0.17(0.02) & -0.10(0.03) & -0.16(0.03) \\
         \bottomrule
    \end{tabularx}
    \caption{Metrics and their standard deviations for predictions of the standard deviation of the subjective ratings by different methods, aggregated over 15 random seeds where applicable. }
    \label{tab:results_uncertainty}
\end{table*}



The results for uncertainty estimation are presented in  Table~\ref{tab:results_uncertainty} and \Cref{fig:std-summary}.
The reported metrics are the coefficient of determination $R^2$, the Pearson correlation coefficients $r_p$, and the Spearman $r_s$ correlation coefficients.
For mean prediction,


The results for mean prediction and uncertainty estimation are presented in Table~\ref{tab:results_mean} and Figure~\ref{fig:mean-summary}. The table also includes the results of TNN-SVR \cite{Cheuk2020RegressionbasedMusicEmotion}, a baseline model for music emotion recognition, as a comparison.\footnote{Note that the TNN-SVR results are computed on the same dataset but using a different data split and may not be directly comparable.} We observe that all of the methods have fairly similar performance on both valence (V) and arousal (A) predictions. Perhaps the only notable difference is the slightly lower performance of the KLD loss in the prediction of the valence ratings. 



The results for uncertainty estimation are presented in  Table~\ref{tab:results_uncertainty} and \Cref{fig:std-summary}. We find that none of the methods capture the ground-truth standard deviation consistently on either valence or arousal. All $R^2$ metrics are effectively well below zero, indicating that the uncertainty estimations of all methods are significantly worse than a naive linear predictor. Both correlation coefficients for all methods are also either weakly negative or close to zero. This indicates that the model generally could not predict the uncertainty estimates with any appreciable correlation to the interrater variations of the ratings. 

By further visualizing the relationship between the empirical and predicted standard deviations in \Cref{fig:std-summary}, however, it can be observed that each method exhibits slightly different ``failure'' behaviors. 
From this, a few conjectures can be made about the limitation of each uncertainty quantification technique employed.

The multiple-seed and MC dropout methods consistently underestimate the empirical standard deviations.
For the multiple-seed method, the variations in the outputs of differently randomized but otherwise similarly trained models appear to not at all reflect the interrater variations. In other words, a different random seed cannot be successfully used to mimic the effect of a different rater. Similarly, the randomly disconnected activations in the MC dropout method  appear to not be an effective method for quantifying the interrater variability.

The model trained on NLL loss also largely predicted standard deviations within a relatively narrow range, but with a few outliers with large predicted standard deviations. Models trained on the MSE and KLD losses, on the other hand, exhibited a larger range of output values, yet the predicted and empirical standard deviations appear nearly uncorrelated.

The different scales of standard deviation prediction can be explained by the loss functions. The multiple-seed and MC dropout methods do not involve the standard deviation in the loss at all, and therefore the scale of standard deviation is completely unknown.
NLL loss, on the other hand, pushes $\hat{\sigma}$ towards zero while it allows larger $\hat{\sigma}$ values when the mean prediction deviates more from the empirical mean.
MSE loss and KLD loss, with the help of empirical standard deviation, enforce the prediction to be on a similar scale to the empirical value.

\section{Discussion}

Overall, our study results indicate that none of the investigated methods can effectively model the uncertainties associated with interrater disagreements.
The reasons for this could be related to data, either due to an insufficient amount (of either audio data, or number of ratings per data point), or due to inherent label noisiness of an ill-defined task \cite{Lerch2023MoodRecognition}.

In practice, even though we are more interested in the data uncertainty caused by the subjectivity of MER, this type of uncertainty can hardly be disentangled from model uncertainty or other types of data uncertainty. While model uncertainty is considered reducible with additional high-quality data, data uncertainty reflects the intrinsic subjectivity or noise in the data. As a result, additional experiments with varying amounts of data might be required to better understand the sensitivity of each method to each type of uncertainty. 

It should perhaps be noted that our negative results are in line with the results of a recently published work in image classification \cite{Mucsanyi2024BenchmarkingUncertaintyDisentanglement}, using a large number of methods, some much more complex than the ones presented here.
Neither do these methods provide accurate uncertainty estimations, nor do they successfully disentangle between model and data uncertainty.
These results, published after we conducted our experiments, indicate that obtaining reliable uncertainty estimates remains difficult even with significant data and resources; it is, therefore, ultimately unsurprising that uncertainties also cannot be reliably estimated in the MER regression task.

Finally, it should be noted that ---~since the foundation model in our setup is pre-trained~--- any random initializations or random dropouts only apply to the FCN portion of the model. While the use of frozen pretrained foundation models with a small FCN on top is not uncommon in UQ \cite[see][Section 3.2]{Kirchhof2023URLRepresentationLearning}, it is unclear how this may have affected the downstream uncertainty estimates. It is conceivable that the information about uncertainties may have been lost in the embedding extraction process within the foundation model, given the random-masking technique utilized in the training of MusicFM \cite{Won2024FoundationModelMusic}.

\section{Conclusion}\label{sec:conclusion}
In this study, we investigated several current methods for uncertainty estimation at the example of music emotion recognition, a highly subjective regression task with considerable annotator disagreements.

More specifically, we test MSE loss and KLD loss, which require empirical uncertainty during training, and NLL loss, different random seeds, and MC dropout, which do not require empirical uncertainty.
Results show that~---~while largely delivering prediction of the mean with expected accuracy~---~the investigated methods fail at estimating the standard deviation of the distribution of annotations, the uncertainty. This is even the case if the ground-truth standard deviation is added as a training target. In some cases, the uncertainty estimation predicts only noise, in other cases, some minor correlation with the target ground truth can be found, however, none of the investigated methods provide sufficiently useful results.

Thus, our results indicate that the investigated methods are insufficient to model uncertainty in data. This means that current systems fail to take into account the inter-rater variability of subjective data and, therefore, this variability cannot be modeled. Future work needs to explore other less common approaches for modeling this uncertainty, as the typically used deterministic machine learning models are apparently unsuited for this task. 

\ifdoubleblind\else
\section{Acknowledgments}

K.~N.~Watcharasupat was supported by the Google PhD Fellowship in Machine Perception (from 2024), the American Association of University Women (AAUW) International Fellowship (2023--2024), and the IEEE Signal Processing Society Scholarships (from 2023).

The authors would like to thank Suhel Keswani for his assistance with the project.

\fi



\bibliographystyle{splncs04}
\bibliography{references,2025-ecir}

%
%
%
%
%

\end{document}